\documentclass[aps,prl,twocolumn,nofootinbib,superscriptaddress,floatfix]{revtex4-1}  
\usepackage{graphicx}  
\usepackage{amsmath}
\usepackage{amssymb}   
\usepackage{color}
\usepackage{hyperref}
\usepackage{ulem}
\usepackage{natbib}
\usepackage{enumitem}

\newcommand{\ltsima} {$\; \buildrel < \over \sim \;$}
\newcommand{\gtsima} {$\; \buildrel > \over \sim \;$}
\newcommand{\lta} {\lower.5ex\hbox{\ltsima}}
\newcommand{\gta} {\lower.5ex\hbox{\gtsima}}

\newcommand{\OneA}{{IA}\ }

\def\ln{\mathrm{ln}}

\newcommand{\C}{{\sf{C}}}

\newcommand{\I}{{\sf{I}}}

\newcommand{\T}{{\sf{T}}}

\begin{document}

\title{Standard rulers, candles, and clocks from the low-redshift Universe}

\author{Alan Heavens}\email{a.heavens@imperial.ac.uk}
\affiliation{Imperial Centre for Inference and Cosmology, Imperial College, Blackett Laboratory, Prince Consort Road, London SW7 2AZ, U.K.}

\author{Raul Jimenez}\email{raul.jimenez@icc.ub.edu}
\affiliation{ICREA \& ICC, University of Barcelona, Mart{\'\i} i Franqu{\`e}s 1, E-08028, Barcelona, Spain}
\affiliation{Institute for Applied Computational Science, Harvard University, MA 02138, USA}

\author{Licia Verde}\email{liciaverde@icc.ub.edu}
\affiliation{ICREA \& ICC, University of Barcelona, Mart{\'\i} i Franqu{\`e}s 1, E-08028, Barcelona, Spain}
\affiliation{Institute of Theoretical Astrophysics, University of Oslo, 0315 Oslo, Norway}

\date{\today}

\begin{abstract}
We measure the length of the Baryon Acoustic Oscillation (BAO) feature, and the expansion rate of the recent Universe, from low-redshift data only, almost model-independently.  We make only the following minimal assumptions:  homogeneity and isotropy; a metric theory of gravity; a smooth expansion history, and the existence of standard candles (supernov\ae) and a standard BAO ruler. The rest is determined by the data, which are compilations of recent BAO and Type \OneA supernova results.    Making only these assumptions, we find for the first time that the standard ruler has length $103.9 \pm 2.3\, h^{-1}$ Mpc.  The value is a {\it measurement}, in contrast to the model-dependent theoretical {\it prediction} determined with model parameters set by Planck data ($99.3 \pm 2.1 \, h^{-1}$ Mpc).  The latter assumes $\Lambda$CDM, and that the ruler is the sound horizon at radiation drag.   Adding passive galaxies as standard clocks or a local Hubble constant measurement allows the absolute BAO scale to be determined ($142.8\pm 3.7$ Mpc), and in the former case the additional information makes the BAO length determination more precise ($101.9\pm 1.9 \, h^{-1}\,$Mpc).  The inverse curvature radius of the Universe is weakly constrained and consistent with zero, independently of the gravity model, provided it is metric.  We find the effective number of relativistic species to be  $N_{\rm eff} = 3.53\pm 0.32$, independent of late-time dark energy or gravity physics.  
\end{abstract}

\maketitle


\section{Introduction}
Standard candles and standard rulers have been instrumental in the development of the cosmological model, with Type \OneA supernovae being used to establish the acceleration of the Universe, and the sound horizon at decoupling being used in conjunction with Baryon Acoustic Oscillations (BAOs) to constrain early Universe physics (see e.g., \cite{DETF, Weinberg2013}).  We can add standard clocks \cite{53w1,53w2} -  objects whose ages are measured independently of the cosmological model, and which were born so early that scatter in formation time is negligible compared to the age of the Universe.  The cosmological importance of the BAO scale is that it is a key theoretical prediction of models, depending on the sound speed and expansion rate of the Universe at early times, before matter and radiation decouple.  In combination with lower redshift measurements this can  be used to constrain, for example, the number of relativistic species  including neutrinos \cite{EisenWhiteUncert}.  

The main purpose of this study is to provide a measurement of the BAO scale, which will survive even if $\Lambda$CDM does not.  It decouples the physics at $z\simeq 0$ from physics at the time when the BAO scale is set (typically $z\gtrsim 1000$) and allows theoretical models to be confronted with the BAO scale independently of assumptions of properties of conventional dark energy.  In variants of the standard model, this means, for example, that our  conclusions about the number of neutrino species rely only on the relatively simple matter- and radiation-dominated physics in the pre-radiation drag era.  Other models can be very simply tested against the BAO measurement provided only that a theoretical prediction of the scale can be made.


The key link between the standard objects is the Hubble parameter, and its dependence on redshift $H(z)=\dot R/R$, where $R$ is the scale factor.  In this paper, we assume simply the existence of standard objects and an expansion rate, and allow low-redshift data from supernovae and galaxy clustering to constrain weakly the curvature of the Universe, without assuming General Relativity\footnote{In reality the adopted BAO scale has in some cases used a reconstruction technique that, assuming Newtonian gravity, lessens the small shift and degradation of the signal due to gravitational evolution, but the difference in position of the peak of the angle-averaged correlation function is relatively small compared with current error bars.}.
 This procedure recovers the expansion history from redshift 0 to 1.3 to a precision of  2.5\% (11\%) at $z=0\ (1.3)$ and provides a weak curvature constraint, but most interestingly, measures the BAO scale independently of the cosmological model as $r_d = 103.9 \pm 2.3\,h^{-1}$ Mpc ($101.9 \pm 1.9\,h^{-1}$ Mpc with clocks).   Given the importance of the BAO scale to cosmology, a measurement independent of all but these very mild assumptions is extremely useful.

Optionally adding standard clocks (passive galaxy ages) or local Hubble parameter measurements allows an absolute BAO scale determination (in Mpc), and clocks add some statistical power. We find excellent agreement with the derived quantity of the sound horizon deduced from Planck data \cite{Planck2013a}, which assumes $\Lambda {\rm CDM}$.   The main difference with other studies that use similar datasets (e.g. \cite{Wang,Mortsell,Antonio}) is that here we {\it measure} the standard ruler length, the expansion history,  and the curvature simultaneously, without cosmological model assumptions beyond weak requirements on symmetry and smoothness.  The CMB-derived BAO scale is completely different - it is a model-dependent theoretical prediction, to be confronted with the measurement presented here.


\section{Theory and assumptions}

Assuming the cosmological principle of homogeneity and isotropy, the metric may be written
\begin{equation}
ds^2 = c^2 dt^2 - R^2(t)\left[dr^2 + S_k^2(r)\left(d\theta^2+\sin^2\theta d\phi^2\right)\right]
\end{equation}
where  symbols have their usual meanings and the scale factor $R(t)$ has the dimensions of length.  The form of the metric assumes only symmetry, and not the gravity model, which is needed to determine $R(t)$.  $S_k(r)=\sin r,r,\sinh r$ depending on the curvature of the Universe $k=1,0,-1$.  $1+z=R_0/R(t)$, where $R_0$ is the present value of the scale factor, and 
\begin{equation}
r(z) = \frac{c}{R_0 H_0} \int_0^z \frac{dz'}{E(z')} \equiv \frac{c}{R_0 H_0} \tilde r(z),
\end{equation}
where $E(z)\equiv H(z)/H_0$.  The angular diameter distance is
\begin{equation}
D_A(z) =  (1+z)^{-1}\frac{c}{H_0 \kappa}S_k\left(\kappa \tilde r\right),  
\end{equation}
where $\kappa \equiv c/(R_0 H_0)$ is the inverse curvature radius in units of $H_0/c$, and the curvature radius for $k=\pm1$ is $ k \,R_0$, and infinite for $k=0$.  For any metric theory of gravity, the luminosity distance is $D_L=(1+z)^2 D_A$.  If we also assume General Relativity, we can identify $\kappa$ with the curvature density parameter, through  $\Omega_k=k \kappa^2$.

Assuming Type \OneA supernovae can be made standard candles (with some absolute magnitude, $M\simeq -19.1$\cite{Betoule2014}), their apparent magnitude $m$ determines the distance modulus
$\mu(z)\equiv m-M =  25+5\log_{10} \left[D_L(z)/{\rm Mpc}\right]$.

For the BAOs (see e.g. \cite{Weinberg2013}),  angle-averaged clustering data determine $D_V(z)/r_d$ where
\begin{equation}
D_V(z) \equiv \left[(1+z)^2D^2_A(z) \frac{cz}{H(z)}\right]^{1/3}.
\end{equation}
$r_d$ is the length of a standard ruler.  For the measurement, we make no assumptions about its origin, but it is normally interpreted as the sound horizon at the end of radiation drag $z_d$, 
\begin{equation}
r_d=\int_{z_d}^{\infty}\frac{c_s(z)}{H(z)}dz,
\end{equation}
where $c_s(z)$ is the sound speed. 

We parametrise\footnote{This choice is motivated by its appearance in the length integrals.} the cosmology by $h^{-1}(z) \equiv 100 {\rm km}\,{\rm s}^{-1}{\rm Mpc}^{-1}/H(z)$, specified at $N\simeq 6$ values equally-spaced in $0<z<1.3$ and linearly-interpolated.    For the supernov\ae, we allow an offset in the absolute magnitude compared with the standard value, $\Delta M$, so we do not assume their luminosity.  Similarly, for BAO measurements, we assume only that there is a standard ruler, parametrised by $r_d \equiv \hat r_d h^{-1}$.  The parameters are therefore $(\hat r_d,\Delta M,\Omega_k,h^{-1}(0), h^{-1}(z_1),\ldots,h^{-1}(z_N))$.  Uniform priors are assumed.

Our main result is based on supernov\ae\ and BAOs alone, but we can add clocks, or a gaussian prior on $h \equiv h(z=0)=0.738\pm 0.024$ \cite{Riess2011}.   For the clocks, we use passive elliptical galaxy ages determined from analysis of stellar populations, and assume that the formation time was sufficiently early that variations in formation time are negligible in comparison with the Hubble time.  Differential ages (see \cite{moresco} for discussion of this method) then give the inverse Hubble parameter, $\delta t(z) \simeq \delta z/[H(z)(1+z)]$.  This adds a little statistical power.  Adding either of these sets an absolute scale, and allows a determination of $r_d$ in Mpc, rather than $h^{-1}\,$Mpc.

\section{Data}

{\bf Supernovae}. We use the compilation \cite{Betoule2014} of 740 Type \OneA supernov\ae\ binned into 31 redshift intervals between $0$ and $1.3$, and their covariance matrix.  The binning and the central limit theorem motivate a gaussian likelihood.  


{\bf BAO.}  The BAO data are measurements of $D_V/r_d$, from 6dF ($z=0.106$) \citep{Beutler2011}, WiggleZ ($z=0.44,\ 0.6$ and $0.73$) \citep{Blake2011,Kazin2014},  and BOSS ($z=0.32, 0.57$)\citep{Tojeiro2014,Anderson2014}.   We use the covariance matrix in \cite{Blake2011} for WiggleZ.

%


%
{\bf Clocks.} 
We combine measurements of \cite{moresco,simon,stern}, giving 16 $H(z)$ measurements  \cite{VPJ2014} in $0.1 < z < 1.3$.


\begin{figure*}
\begin{center}
\includegraphics[width=2\columnwidth]{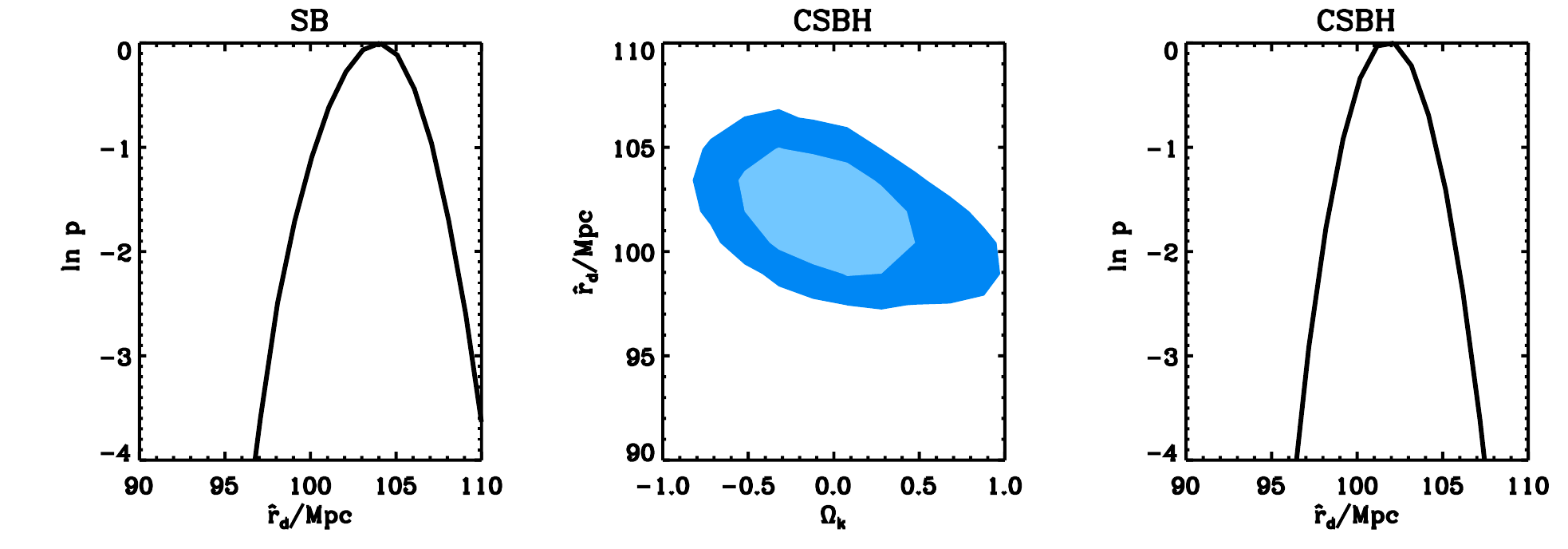}
\end{center}
\caption{Unnormalised  probabilities for $r_d/h^{-1}{\rm Mpc}$ from supernov\ae\ and BAOs (left panel) and for $\hat r_d$ and curvature (expressed as a GR equivalent $\Omega_k$) with clocks and a Hubble prior (centre and right).}
\label{FigChain}
\end{figure*}

\section{Results}

The posterior probability of the parameters is obtained from the likelihood
\begin{eqnarray}
2\ln L &=& {\rm constant} - \sum_{i=1}^{16} \frac{\left[H(z_i)-H_i\right]^2}{\sigma_{Hi}^2} \\\nonumber 
& -& \sum_{i,j=1}^{31}\left[\mu(z_i)-\mu_{i}\right](\C_{\rm SN})^{-1}_{ij}\left[\mu(z_j)-\mu_{j}\right] \\\nonumber 
& -& \sum_{i,j=1}^{6}\left[D_V(z_i)-D_{Vi}\right](\C_{\rm BAO})^{-1}_{ij}\left[D_V(z_j)-D_{Vj}\right],
\end{eqnarray}
multiplied optionally by the $h$ prior \citep{Riess2011}.  We run MCMC chains of $10^7$ points, removing a burn-in of $10^6$ points, and thinning by a factor 10.   A Gelman-Rubin test shows good convergence, with parameter $R=1+O(10^{-4})$.  We find no evidence for tension between the three datasets, with the parallel expansion rate $H_{\parallel}$, determined by the $t-z$ relation, being consistent with the supernov\ae\ and BAOs with the same $H(z)$.

\begin{figure}
\begin{center}
\includegraphics[width=\columnwidth]{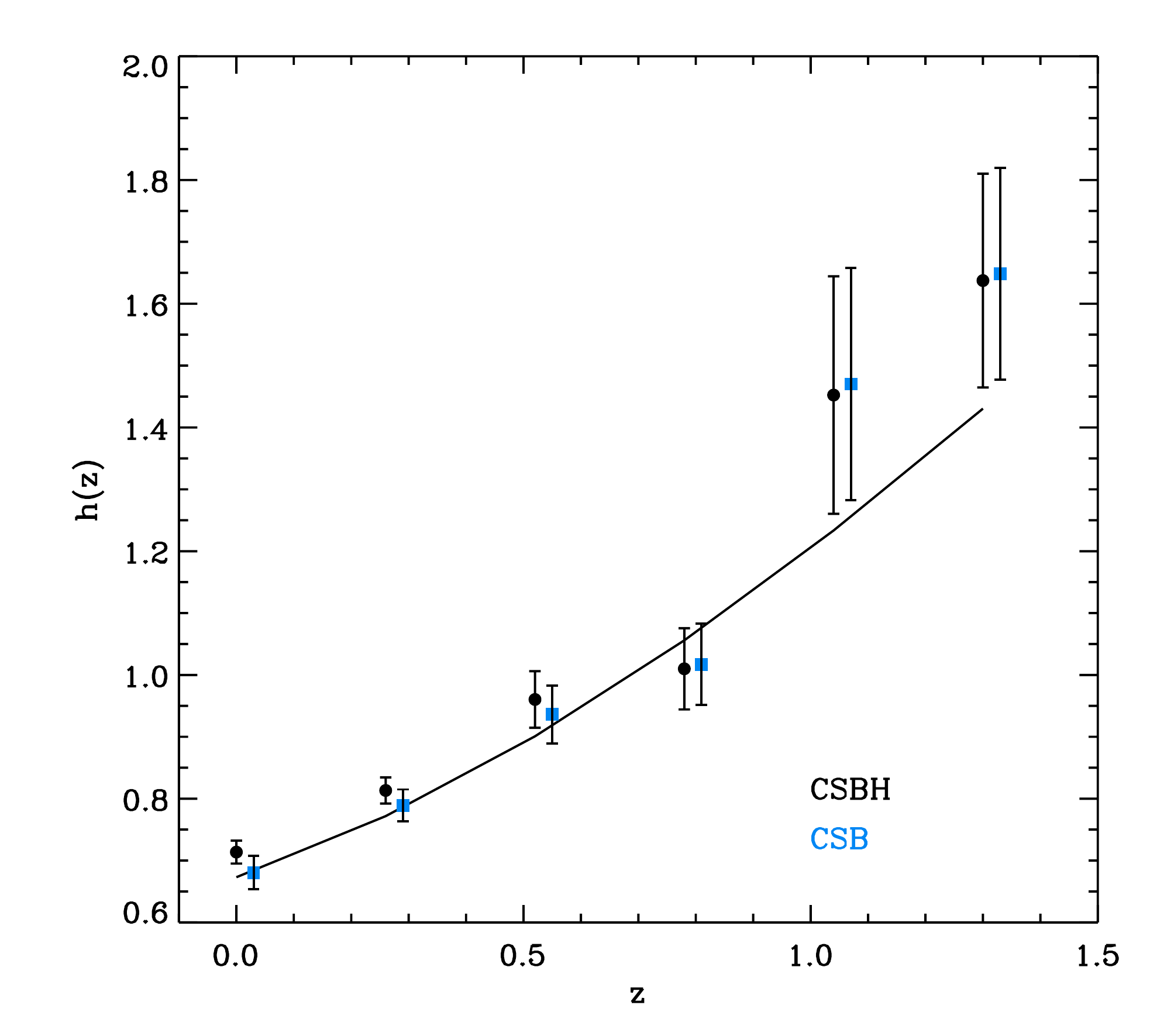}
\end{center}
\caption{The recent expansion rate of the Universe determined from clocks, supernov\ae, and BAOs, with (CSBH; black circles) and without (CSB; blue squares) a prior on the present Hubble parameter.  Squares have been offset in redshift for clarity.  We also show the Planck best-fit $\Lambda {\rm CDM}$ expansion history with $\Omega_m=0.315$, $\Omega_\Lambda=0.685$ and $h=0.673$ \citep{Planck2013a}.}
\label{Hz}
\end{figure}

 Fig.~\ref{FigChain} shows the posteriors for $\hat r_d$ and $\Omega_k$ for supernov\ae\ and BAOs (left) and with clocks and a Hubble prior added (centre, right). Fig. \ref{Hz} shows the derived expansion history.  Without a Hubble prior, $h$ is inferred to be $0.68\pm 0.03$. In Table \ref{Results} we summarise the marginal posteriors.

\begin{table*}
\begin{tabular}{lccccccccc}
\hline
Data & $r_d/h^{-1}{\rm Mpc}$ &  $\Delta M$ &  $k (c/H_0 R_0)^2\ [=\Omega_k]$ & $h^{-1}$ & $h^{-1}_{0.26}$ & $h^{-1}_{0.52}$ & $h^{-1}_{0.78}$ & $h^{-1}_{1.04}$ & $h^{-1}_{1.3}$  \\
\hline
SBH		 & $103.9 \pm 2.3 $ & $ 0.10 \pm 0.08 $ & $ -0.78 \pm 0.48 $ & $ 1.37 \pm 0.05 $ & $ 1.22 \pm 0.05 $ & $ 1.10 \pm 0.07 $ & $ 1.03 \pm 0.13 $ & $ 2.05 \pm 0.98 $ & $ 3.39 \pm 6.39 $\\
CSB 		 & $ 100.7 \pm 2.0 $ & $ -0.06 \pm 0.08 $ & $ 0.36 \pm 0.42 $ & $ 1.47 \pm 0.06 $ & $ 1.27 \pm 0.04 $ & $ 1.07 \pm 0.05 $ & $ 0.99 \pm 0.06 $ & $ 0.69 \pm 0.09 $ & $ 0.61 \pm 0.07$ \\
CSH 	 & $ - $ & $ 0.04 \pm 0.06 $ & $ 0.09 \pm 0.38 $ & $ 1.40 \pm 0.04 $ & $ 1.23 \pm 0.03 $ & $ 1.02 \pm 0.07 $ & $ 1.01 \pm 0.07 $ & $ 0.69 \pm 0.09 $ & $ 0.62 \pm 0.07 $\\
CBH		 & $ 107.1 \pm 4.6 $ & $ - $ & $ 0.12 \pm 1.3 $ & $ 1.37 \pm 0.04 $ & $ 1.29 \pm 0.06 $ & $ 1.07 \pm 0.07 $ & $ 0.99 \pm 0.07 $ & $ 0.67 \pm 0.09 $ & $ 0.61 \pm 0.07 $\\ 
CSBH	 & $  101.9 \pm 1.9 $ & $ 0.04 \pm 0.06 $ & $ 0.06 \pm 0.37 $ & $ 1.40 \pm 0.04 $ & $ 1.23 \pm 0.03 $ & $ 1.04 \pm 0.05 $ & $ 1.00 \pm 0.06 $ & $ 0.70 \pm 0.10 $ & $ 0.62 \pm 0.07 $\\
\hline
\end{tabular}
\caption{Posterior mean and standard deviation for the model parameters.  The parameters $h^{-1}_z$  are labelled by $z$, except for  $h^{-1}$ which is $z=0$.  CSBH refer to clocks, supernov\ae, BAOs and Hubble prior. Without clocks, the high-$z$ expansion rate is poorly-constrained. Dropping the Hubble prior from line 1 does not alter $r_d/h^{-1}{\rm Mpc}$ at all.}
\label{Results}
\end{table*}

\section{Discussion}

\begin{figure}
\begin{center}
\includegraphics[width=0.35\columnwidth]{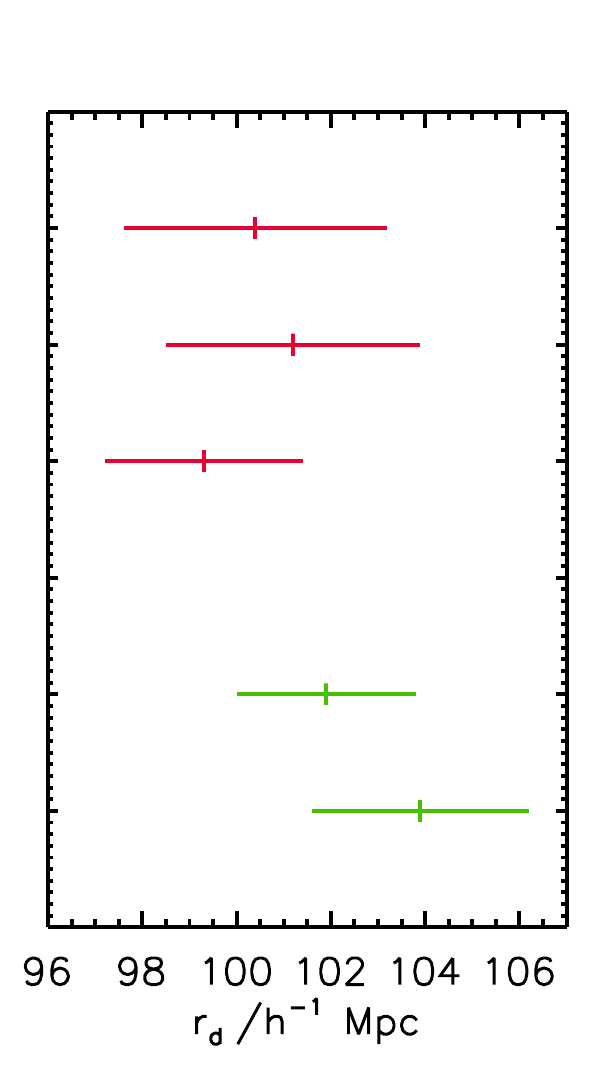}
\includegraphics[width=0.52\columnwidth]{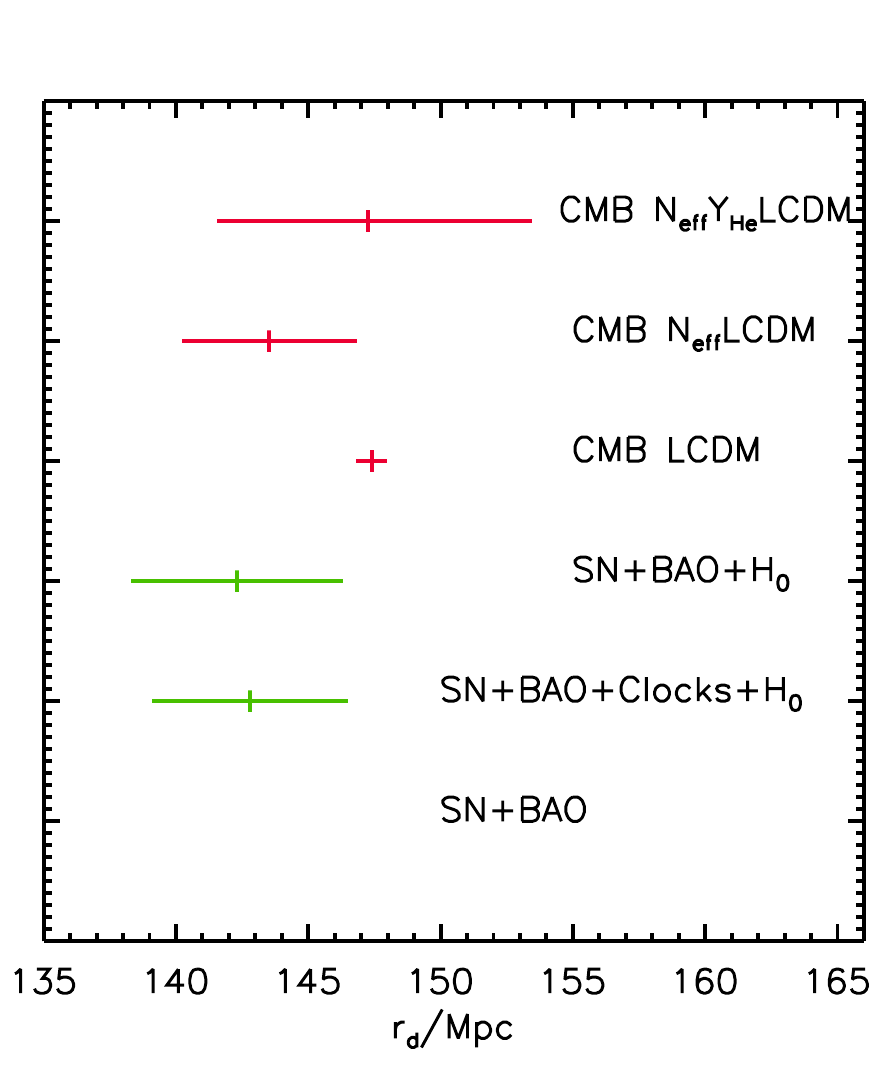}
\end{center}
\caption{Cosmological model-independent BAO length measurement from different low-redshift dataset combinations (bottom; green) compared with theoretical predictions assuming $\Lambda$CDM and extensions (red; top), where errors reflect uncertainties in model parameters from Planck.  Right panel shows BAO lengths in Mpc, left panel in $h^{-1}\,$Mpc. See text for more details.}
\label{figrdhat}
\end{figure}

We have measured the length of the BAO scale 
and determined the expansion history of the recent Universe in an almost model-independent way, using supernov\ae\ and BAO measurements, with and without passive galaxy clocks and a prior on the current value of the Hubble parameter.  We assume only homogeneity and isotropy, a metric theory of gravity, a smooth expansion history, and the existence of standard rulers and candles; the rest is determined by the data.   Using a compilation of supernova data \citep{Betoule2014} and Baryon Acoustic Oscillation measurements  \citep{Beutler2011,Kazin2014,Tojeiro2014, Anderson2014},  we determine for the first time a precise measurement of the standard ruler length ${r}_d = 103.9\pm 2.3\, h^{-1}$ Mpc, and adding clocks shifts the peak and reduces the error slightly, $101.9\pm 1.9\,h^{-1}$ Mpc.  With clocks and a Hubble prior, this can be translated into a physical length, $r_d = 142.8 \pm 3.7$ Mpc.  This is in excellent agreement with the model-dependent theoretical expectation, with parameters determined from the CMB, for which Planck publicly-available MCMC chains give ${r}_d= 99.3\pm 2.1$ $h^{-1}$Mpc   ($147.49 \pm 0.59$ Mpc) for $\Lambda$CDM \citep{Planck2013a},  assuming the ruler is the sound horizon at radiation drag ($z\simeq1059$).   Extending the $\Lambda$CDM model to vary the number of relativistic species (e.g., neutrinos) the CMB gives $101.2\pm 2.7\,h^{-1}$ Mpc   ($143.53 \pm 3.3$ Mpc), and allowing also the Helium yield to vary gives $100.4\pm 2.8\,h^{-1}$ Mpc (fig.~\ref{figrdhat})  ($147.25 ^{+6.2}_{-5.7}$ Mpc).   We also find that the other data do not pull the supernova luminosity away from its value determined internally. We also obtain weak, but model-independent, constraints on curvature. 
%

\begin{figure}
\begin{center}
\includegraphics[width=\columnwidth]{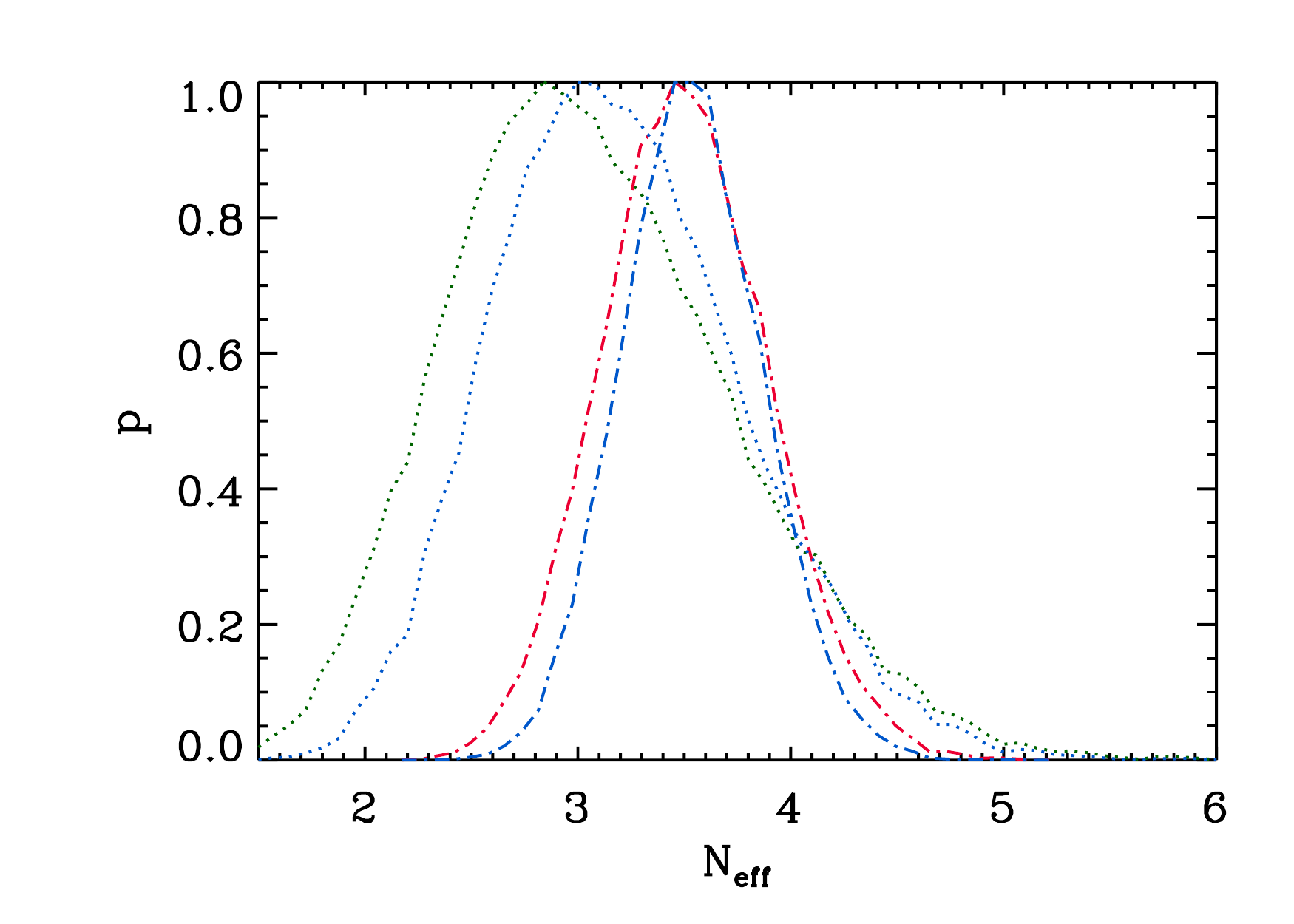}
\end{center}
\caption{The posterior for the effective number of relativistic species in the early Universe, in an extended $\Lambda$CDM model using Planck likelihood chains (dot-dash), and allowing the Helium yield to vary (dotted).  Blue curves (peaking at slightly higher $N_{\rm eff}$) use the $\hat r_d$ and its error, and do not depend on the properties of dark energy, provided that it is negligible at $z>1000$.}
\label{Neff}
\end{figure}

Our results are insensitive to how we parametrise the expansion history; setting $N=5$ or $7$, or interpolating in $h$, gives the same $r_d$ and $h(z)$ to within a very small fraction of the statistical error.  The results are fairly insensitive to the inclusion or exclusion of different datasets, with a small ($<2\%$) decrease in $r_d$ if clocks are included in the analysis.   

Normally, a cosmological model such as $\Lambda$CDM is assumed from the Big Bang to the present day, and data, especially from $z\simeq 10^9$ (nucleosynthesis), $z\simeq 10^3$ (recombination) and $z\simeq 0$ are used to confront the model.  This cradle-to-the-grave approach is an attractive application of the scientific method, but by determining the BAO scale independently of the cosmological model, we are able to isolate near-recombination physics from late-time physics. In doing so we avoid parameters (such as $N_{\rm eff}$) being pulled away from their correct values by an incorrect model trying to fit the low-redshift data.

If we assume $r_d$ is the sound horizon, the low-$z$ measurements limit the scope of new physics to alter the early expansion rate and sound speed - the early Universe physics have to give this BAO length, regardless of what happens at late times.  The conclusions are independent of assumptions of late-time physics since the CMB can predict $r_d$ independently of late dark energy:  odd and even peak heights and Silk damping fix the baryon-to-photon ratio, and the amplitudes of the peaks fix the ratio of matter to radiation density\cite{EisenWhiteUncert}.  By importance-sampling the Planck chains that vary the effective number of relativistic species $N_{\rm eff}$, we obtain $N_{\rm eff} = 3.53\pm 0.32$, which compares with $3.45\pm 0.36$ from $\Lambda$CDM + Planck, but our analysis only assumes that the early dynamics are driven by matter and radiation, and late dark energy is irrelevant. Varying the Helium yield changes $N_{\rm eff}$ from $2.84^{+0.80}_{-0.48}$ to $3.00^{+0.72}_{-0.48}$ (fig.~\ref{Neff}).   Allowing this variation neatly decouples the $z\simeq 10^3$ physics from the $z\simeq 10^9$ physics as well as from the $z\simeq 0$ physics - very different epochs, so the conclusions are robust both to changes in very early Universe physics and late-time physics.

Finally, we note that with precise measurements of $\hat r_d$, we might hope to detect evolving late-time distortions in the observed ruler length due to redshift distortions and nonlinear effects.  However, including a linear gradient of $r_d$ with redshift gives a null result of $0\pm8$ Mpc/(unit z), but this may be an interesting future investigation.

\paragraph{Acknowledgments.---}
We thank Bruce Bassett, Antonio Cuesta and Roy Maartens for useful discussions.  RJ and LV acknowledge support from Royal Society grant IE140357 and   Mineco grant FPA2011-29678- C02-02.
LV's research is supported by the European Research Council under the European Community's Seventh Framework Programme FP7- IDEAS-Phys.LSS 240117. 
Funding for SDSS-III has been provided by the Alfred P. Sloan Foundation, the Participating Institutions, the National Science Foundation, and the U.S. Department of Energy Office of Science. The SDSS-III web site is http://www.sdss3.org/. Some of the results presented rely on observations obtained with Planck (http://www.esa.int/Planck), an ESA science mission with instruments and contributions directly funded by ESA Member States, NASA, and Canada.


\end{document}